# Continuous hydrothermal flow synthesis of Gd-doped CeO$_2$ (GDC) nanoparticles for inkjet printing of SOFC electrolytes

Yu Xu[1], Nicholas Farandos[2], Massimo Rosa[1], Philipp Zielke[1], Vincenzo Esposito[1], Peter Vang Hendriksen[1], Søren Højgaard Jensen[1], Tao Li[2], Geoffrey Kelsall[2], Ragnar Kiebach[1]

[1]Department of Energy Conversion and Storage, Technical University of Denmark, Risø Campus, Roskilde, Denmark
[2]Department of Chemical Engineering, Imperial College London, London, UK

Correspondence author, Ragnar Kiebach, Email: woki@dtu.dk

ABSTRACT
Gd$_x$Ce$_{1-x}$O$_{2-d}$ (GDC) nanoparticles were synthesized, using continuous hydrothermal flow synthesis. By varying the synthesis conditions, particle size and morphology could be tailored. Here, particle sizes between 6 and 40 nm with polyhedral or octahedral shape could be obtained. Gd$_{0.2}$Ce$_{0.8}$O$_{2-d}$ nanoparticles were further processed into inks for inkjet printing. Despite the small particle size/large surface area, inks with excellent printing behavior were formulated. For proof-of-concept, thin GDC layers were printed on a) green NiO-GDC substrates, and on b) presintered NiO-YSZ substrates. While no dense layers could be obtained on the green NiO-GDC substrates, GDC nanoparticles printed on NiO-YSZ substrates formed a dense continuous layer after firing.

## INTRODUCTION

Ceria-based oxides are attractive for several technological applications. For instance, they are applied as electrolytes in solid oxide fuel cells (SOFCs)[1,2] and as ionic conductors in dual-phase oxygen transport membranes (OTMs).[3,4] CeO$_2$ also can be employed as a catalyst or catalyst support for noble metals or oxides applied for catalytic gas oxidation reactions.[5-7]

Among the possible synthesis routes, the solvothermal/ hydrothermal synthesis of CeO$_2$-based oxides is often used when aiming for nanosized materials.[8] Usually taking place in an aqueous system, hydrothermal synthesis of CeO$_2$-based oxide nanomaterials proceeds by hydration of metal salts followed by dehydration, crystallization, and growth of oxides from a supersaturated solution at an increased temperature and pressure. The size and morphology of CeO$_2$ nanoparticles can be tailored by adjusting process parameters such as pH value, reaction time, counter anions and addition of surfactants.[9-12]

In recent years, remarkable progress has been made in pushing forward the hydrothermal production of nanomaterials to a large industrial scale. This is mainly achieved by applying continuous hydrothermal flow synthesis (CHFS). In CHFS, a flow-type apparatus (CHFS reactor), instead of a fixed-volume batch autoclave found in the conventional hydrothermal synthesis, is used.[13,14] One of the essential parts of a CHFS reactor is a mixer, where a precursor flow constantly fed at ambient temperature is mixed with a supercritical water flow (scH2O, T ≥ 374°C, P ≥ 22.1 MPa). As a result, the precursor is rapidly heated up by the scH2O to a (near-) supercritical state, and a solution with a high degree of supersaturation is generated in a short time. As the driving force (i.e. the degree of supersaturation) is high, nuclei are formed almost instantly upon mixing. Therefore, the $scH_2O$ provides an ideal environment for hydrothermal synthesis especially of nanomaterials that can be conducted in a continuous flow way.[15] Various types of nanomaterials prepared by CHFS including oxides,[16,17] sulfides,[18] metals,[19] and metal organic frameworks[20] have been reported. Moreover, the scalability of the CHFS to an industrial-level production has been tested in a pilot plant,[21] a commercial plant (Hanwha Chemical, Republic of Korea),[22] and in the European FP7 project SHYMAN (www.shyman.eu).[20] The syntheses of $CeO_2$ nanoparticles, nondoped[23,24] and doped with La, Pr,[24] Bi[25] or Zn[26] by CHFS have been reported in literatures. Especially the preparation of $Zr_xCe_{1-x}O_2$ nanoparticles by CHFS with a controlled composition attracted attention since this material has a high oxygen storage capacity and is used in three-way catalysts for vehicle exhaust treatment.[27-29]

Another advantage of CHFS is that the nanoparticles are prepared in an 'all-wet' flow process in water, and organic modifiers can be introduced in situ shortly after the formation of nanoparticles, which facilitates tuning the colloidal stability of nanoparticle suspensions[30], especially in water. The possibility of producing large amounts of nanoparticle suspensions in a sustainable, reproducible manner opens up the opportunity to use these nanoparticle suspensions as feedstock/basis for inks suitable for inkjet printing.[31] This additive manufacturing technique offers the potential to fabricate the functional layers of SOFCs with high reproducibility and high level of customization. Here, the fabrication of thin electrolytes[31] which would increase the cell performance of structured electrodes is of particular interest. Inkjet printing is particularly suited to fabricate thin films from colloidal particle dispersions (e.g. $Y_{0.16}Zr_{0.84}O_{2-d}$,[31-33] $Gd_{0.1}Ce_{0.9}O_{2-d}$[34] $La_{0.6}Sr_{0.4}Co_{0.2}Fe_{0.8}O_{3-d}$ - $Gd_{0.1}Ce_{0.9}O_{2-d}$[35]), preferably aqueous dispersions, which are environmentally friendly and suitable for a large-scale industrial deployment.

Here, the one-step continuous production of $Gd_xCe_{1-x}O_{2-d}$ nanoparticles is reported. The influence of pH on the morphology, composition, and size of the nanoparticles was investigated. Moreover, the combination of CHFS and inkjet printing, two highly scalable techniques, was demonstrated. A scaled-up CHFS of $Gd_{0.2}Ce_{0.8}O_{2-d}$ nanoparticles was conducted with increased precursor concentrations to obtain nanoparticles in amounts suitable for inkjet printing. From the as-prepared wet $Gd_{0.2}Ce_{0.8}O_{2-d}$ nanoparticles, inks were formulated and used to fabricate printed SOFC electrolytes. Importantly, firing of dense GDC electrolytes typically requires sintering temperatures above approximately 1450°C.[36,37] As will be shown, by utilizing GDC nanoparticles, it was possible to obtain a dense GDC layer on top of an NiO-YSZ substrate when firing at 1300°C, i.e., at a firing temperature much more applicable to conventional SOFC manufacturing. A reduction in GDC firing temperature can also be obtained by addition of sintering additives whereas this may be associated with a loss in grain boundary/bulk conductivity[36,38]. Interestingly, the formed GDC layer entailed anchoring points for subsequent air electrode deposition, e.g., SOFC cathode material could be (screen) printed on or infiltrated into this GDC layer (not presented in this paper). This could improve the adhesion and in turn the strength of the air electrode-electrolyte interface, which is known to be a weak interface of conventional SOFCs.[39]

# EXPERIMENTAL

## Preparation of $Gd_xCe_{1-x}O_{2-d}$ by CHFS

$Ce(NO_3)_3 \cdot 6H_2O$ (99%, Sigma Aldrich, St. Louis, MO) and $Gd(NO_3)_3 \cdot 6H_2O$ (99.9%, Sigma Aldrich) were dissolved in deionized $H_2O$ (DI $H_2O$) with concentrations of 0.04 mol $L^{-1}$ for Ce(III) and 0.01 mol $L^{-1}$ for Gd(III). For the up-scaled synthesis, a solution of 0.08 mol $L^{-1}$ Ce(III) and 0.02 mol $L^{-1}$ Gd(III) was used. KOH pellets (≥85%, Sigma Aldrich) were dissolved in DI H2O to make solutions with varying concentrations of 0.1, 0.15, 0.185, 0.3, 1.0 mol $L^{-1}$.

An in-house developed two-stage continuous flow-type apparatus (CHFS reactor) was used to conduct the synthesis, details of which were reported previously.[40] Briefly, two mixers (stages) are integrated into a series within the reactor. The first mixer with a co-flow pipe-in-pipe geometry was used to mix the $scH_2O$ flow fed by a 1/16″ capillary pipe with the precursor flow fed from two side arms of a 1/4″ encapsulating pipe. The precursor flow was a room-temperature premixture of the Ce, Gd nitrates solution (reactant solution), and the KOH solution. Each solution was fed into the reactor by separate pumps, and they were mixed at a T-junction upstream of the first mixer. The outflow of the first mixer was fed subsequently to the second mixer with a countercurrent geometry, and was mixed with an upward inlet flow of DI $H_2O$ at room temperature aiming to mitigate sedimentation. The second mixer can for the synthesis of binary systems was used to feed the precursor of a second material.40 The outflow of the second mixer passed subsequently through a reheater. In this work, however, the reheater was not used. The outflow of the reheater was rapidly cooled down to the room temperature by a water-cooled tube-in-tube heat exchanger. After passing an in-line filter (Swagelok, 90-µm pore size), the product flow was depressurized to atmospheric pressure by a backpressure regulator (Tescom, 26-1700 Series). Three thermocouples were inserted into the reactor, one to measure the temperature (T1) of the feeding $scH_2O$ flow, one to measure the temperature (T2) of the outflow of the second mixer and the other to measure the temperature (T3) of the outflow after the reheater. During synthesis, T1, $T_2$ and $T_3$ were maintained at 396°C, 290°C, and 285°C, respectively. The estimated residence time $(t)^{41}$ was ~29 s (see details on the calculation in Supplementary information). The synthesized particles were collected as colloidal slurries at the reactor outlet. A schematic description of the two-stage CHFS reactor is presented in Figure 1. Synthesis conditions are summarized in Table 1. The particles were separated from slurries by centrifugation, and then washed by DI $H_2O$ and dried for subsequent characterization.

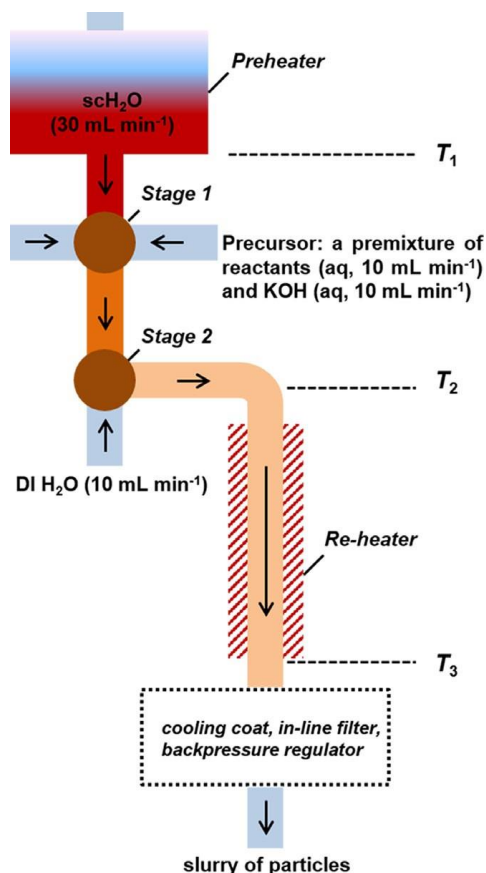

FIGURE 1: A schematic representation of the two-stage CHFS reactor. Pressure was controlled at 26 MPa. The reheater was inactive in this study. The flow rate of each DI H$_2$O/solutions stream is given in parentheses; arrows indicate the flow direction.

Identical conditions as the previous syntheses were applied, except 0.08 mol L$^{-1}$ Ce(III), 0.02 mol L$^{-1}$ Gd(III) nitrates and 1.0 mol L$^{-1}$ KOH solutions were used during the up-scaled CHFS of Gd$_{0.2}$Ce$_{0.8}$O$_{2-d}$ (CG-s in Table 1), to get sufficient amounts of particles for inkjet printing. The conditions were chosen to make sure that Gd$_{0.2}$Ce$_{0.8}$O$_{2-d}$ nanoparticles with the desired composition were obtained. The particles were separated from the slurry by centrifugation and washed with DI H$_2$O.

**Preparation of the primary Gd$_{0.2}$Ce$_{0.8}$O$_{2-d}$ dispersion; ink formulation, printing, and sintering**

**Primary dispersion**

Dispex A40 (Ciba-BASF, UK) was used as dispersant. To prepare the primary dispersion, the wet Gd$_{0.2}$Ce$_{0.8}$O$_{2-d}$ particles (CG-s in Table 1) were redispersed in 50 mL DI H$_2$O and ~0.6 mL Dispex A40 (twelve drops by a 10-mL disposable plastic Pasteur pipette) using an ultrasonic processor (Hielscher UP200St, Germany). The mass fraction of Gd$_{0.2}$Ce$_{0.8}$O$_{2-d}$ was determined by thermogravimetric analysis (TGA, TG 209 F1 Libra, Netzsch, Germany). The viscosity was measured with a rheometer (MCR 302, Anton Paar). The rheological measurements were carried out in rotational mode with a plate-plate configuration at a shear rate of 0-1000 s$^{-1}$. The primary dispersion was separated to two identical batches.

**Printing on green NiO-GDC substrates**

To one batch, polyethylene glycol with a molecular weight 35 000 (PEG 35000) (Sigma-Aldrich) was added at a concentration of 10 mg cm$^{-3}$ to raise the viscosity, and Natsurf 265 added at a concentration of 0.2 mg cm$^{-3}$ to reduce the surface tension to the recommended range for the DMC 10 pL printhead (Fujifilm, Santa Clara, CA)[42] that was used in this study. The surface tension was measured using a tensiometer (Model 250-U1, Rame-hart, SuccaSunna, NJ), and the viscosity was measured using a concentric cylinder viscometer (DV-E Viscometer, Brookhaven) at a shear rate of 2500-12 500 s$^{-1}$. Prior to printing, the ink was first sonicated for 5 minutes (Q55, QSonica, Newtown, CT), and then filtered to remove large aggregates (800 nm syringe filter, Cole Parmer, Hanwell, London, UK).

Three sets of films with 1, 5, 10, and 20 layers were printed onto a green NiO-GDC substrate to make squares of 7 mm side length with 4 minutes intervals between each deposition to allow drying. A Ceradrop X-Serie piezoelectric inkjet printer (Ceradrop, France) was used. The voltage profile (amplitude x, time coordinate y) applied to the piezoelectric printheads was (10 V, 0 μs), (0 V, 5 μs), (0 V, 10 μs), (50 V, 15 μs), (50 V, 20 μs), (10 V, 25 μs).

**TABLE 1** CHFS of Gd-doped CeO$_2$ nanoparticles. Process parameters and microstructural characteristics of the synthesized particles

| Sample | Me$^{3+}$/mol L$^{-1}$ | KOH/mol L$^{-1}$ | pH of slurry$^a$ | Crystallite size/nm (111) | (200) | (220) | Particle size/nm$^b$ | Morphology |
|---|---|---|---|---|---|---|---|---|
| CG1 | 0.05 | 0 | 1.9 | 28.9 | 30.3 | 26.1 | 45(4)$^c$, 39(8)$^d$ | Octahedral |
| CG2 | 0.05 | 0.1 | 3.5 | 8.9 | 9.5 | 8.8 | 8(1) | Polyhedral, spherical |
| CG3 | 0.05 | 0.15 | 5.5 | 6.0 | 6.8 | 5.6 | 6(1) | Polyhedral, spherical |
| CG4 | 0.05 | 0.185 | 9 | 7.2 | 7.5 | 7.0 | 7(1) | Polyhedral, spherical |
| CG5 | 0.05 | 0.3 | 11.9 | 8.2 | 8.4 | 8.0 | 8(2) | Spherical, polyhedral |
| CG6 | 0.05 | 1.0 | 13.1 | 9.2 | 9.9 | 9.0 | 8(2) | Spherical |
| CG-s | 0.1 | 1.0 | 12.8 | 11.9 | 11.9 | 11.5 | 11(2) | Spherical |

Me$^{3+}$ are Ce (III) and Gd(III) cations in the reactant solution; $^a$is the pH of the slurry of nanoparticles; $^b$is an average value of sizes of 80-100 nanoparticles measured in BF-TEM images; $^c$is the average edge size measured with square projections; $^d$is the average edge size measured with rhombic projections. The numbers in the parentheses represent the associated standard deviations.

With a printhead speed of 100 mm s$^{-1}$ across the substrate, the resulting jetting frequency was 3.48 kHz. This resulted in consistent, spherical, droplet ejection, verified with the in-built camera. The 'splat' diameter was measured using the in-built optical microscope in the printer as 80 μm. The diameter overlap for the printed films was 50 μm with a square droplet deposition lattice (i.e. 62.5%). The substrate was prepared by mixing the GDC and NiO powders (GDC and NiO powders supplied by Fuel Cell Materials, Columbus, OH, and Inframat Advanced materials, Manchester, CT, respectively) with dimethyl sulfoxide (VWR) and polyethersulfone (Ameco Performance) and milled for 48 hours (JARMILL, Gladstone Engineering Milton, UK). Polyethylene glycol 30-dipolyhydroxys-tearate (Arlacel P135, Uniqema) was added and milling continued for a further 48 hours. The final composition contained the above species in the weight ratio 60:90:100:20:2, respectively. The suspension was degassed under vacuum, cast (height 0.5 mm) onto a plate, and finally submerged into an external coagulant bath to complete the phase inversion process. The substrate was dried and flattened for 24 hours before printing. After printing, samples were sintered with the following profiles: 0-650°C at 4°C min$^{-1}$; dwell at 650°C for 4 hours (to pyrolyze organics); 650-1300°C at 15°C min$^{-1}$; dwell for 8 hours; cool to room temperature at 4°C min$^{-1}$.

**Printing on presintered NiO-YSZ substrates**

To the other batch, Natsurf 265 was added under magnetic stirring at a concentration of 6 mg cm$^{-3}$ to reduce surface tension to the recommended range for the DMC 10 pL printhead (Fujifilm)[42].

The surface tension of the ink was assessed using a bubble pressure tensiometer (BP 50, Kruss) and the viscosity was measured with a rheometer (MCR 302, Anton Paar). The rheological measurements were carried out in rotational mode with a plate-plate configuration at a shear rate of 0-1000 s$^{-1}$. The plate-plate configuration, though commonly applied for the characterization of (primary) dispersions with higher viscosities, has been successfully used in the past at DTU Energy[32].

A Pixdro LP50 printer equipped with a DMC 10-pL printhead (Fujifilm)[42] was used for depositing the ink. Before printing, the ink was filtered using a syringe filter with a 700 nm mesh. Each sample consisted of 10 layers printed only after complete drying of the material previously printed. The substrate coverage and the optimal splat overlap was achieved by printing with a resolution of 700 DPI (dots per inch). The jetting pulse was optimized in order to jet single spherical droplets and the resulting voltage profile was (0 V, 0 μs), (50 V, 6 μs), (50 V, 16 μs). A jetting frequency of 1 kHz was used.

The substrate[43] was a presintered NiO-YSZ cermet anode on a NiO-YSZ support. The green NiO-YSZ anode was prepared by tape casting (after reduction in NiO, Ni/ YSZ ratio of 40/60 vol/vol, 8 mol% $Y_2O_3$-stabilized $ZrO_2$). Two layers were co-laminated in the green state at ca. 150°C, i.e. a 10-15 μm thick NiO-YSZ anode on top of a 300 μm thick NiO-YSZ support. The firing of the substrate consisted of a multistep debinding procedure up to 700°C for 48 hours in total. The final sintering temperature, 1300°C, was kept for 6 hours.

The printed samples were sintered in air with the following profile: 0-600°C at 0.25°C min$^{-1}$; dwell for 4 hours; 600°C-$T_{max}$ at 1°C min$^{-1}$; dwell for 6 hours; cool to room temperature at 1.67°C min$^{-1}$. Here, three different temperatures $T_{max}$ were used, 800, 1000, and 1300°C.

One may notice that the two groups of inkjet printing of GDC nanoparticles in this study differ in several aspects. For instance, different substrates (green vs presintered, prepared by phase inversion vs tape casting, and NiO-GDC vs NiO-YSZ) were chosen, which affects the printing parameters that were optimized independently in the laboratory. This choice was made deliberately to demonstrate the versatility of the inks made of GDC nanoparticles obtained by CHFS in the inkjet printing process, but made it difficult to compare directly the printing processes. Nevertheless, the same printhead was used, which is a very important detail despite that different printers were applied. In addition, the solid loadings (amount of GDC nanoparticles per unit area) of the inks could be compared and were 0.09 g cm$^{-2}$ on the green NiO-GDC substrates and 0.06 g cm$^{-2}$ on the presintered NiO-YSZ substrates, separately.

## Characterization

### Particle characterization

The pH values of the supernatants of the slurries in Table 1 were measured by a 781 pH/Ion Meter (Metrohm AG, Switzerland) and pH indicator strips (Sigma Aldrich) after the particles were separated. X-ray diffraction (XRD) analysis was conducted on the dried particles by a Bruker Robot D8 diffractometer (Cu Ka radiation, 0.154 nm; 2h 10-120° with a step size of 0.01°). Using the Diffrac Eva Suite (Bruker, Germany), the volume-weighted crystallite size was estimated from the full width at half maximum (FWHM) of the diffraction peaks in the XRD patterns. For TEM characterization, the particles were redispersed in ethanol by means of ultrasonic treatment and dropped onto a holey carbon film/Cu grid. Bright field TEM (BF-TEM) and high-resolution TEM (HR-TEM) images were recorded with a JEOL 3000F microscope operating at 300 kV with a field emission gun. The images were analyzed by the Gatan DigitalMicrograph software (Gatan, Pleasanton, CA) and live fast Fourier transform (FFT) of high-resolution images was performed. To quantify the size of particles, the ImageJ software was used to outline the perimeter of the particles' projections in BF-TEM images, and by assuming a spherical shape, the particle diameters were calculated. To analyze the chemical composition of the particles, they were loaded on carbon tapes and a Hitachi Tabletop TM3000 ('Analysis', 15 kV charge-reduction mode) equipped with a silicon drift detector (SDD, energy resolution 0.154 keV) was used to do a standardless quantification EDS analysis.

The particle surface area (CG-s, particles for making inks) was measured by $N_2$ adsorption employing the Brauner-Emmett-Teller (BET) theory using a Micrometrics 3Flex BET instrument (Canada) after drying the primary dispersion and degassing at 200°C.

### SEM and EDS investigations of the printed samples

The green printed samples on green NiO-GDC substrates were imaged optically (VHX-900 Digital Microscope, Keyence, UK). Sintered samples were broken, and the fracture surfaces were investigated, using a SEM (TM3030 Table top Microscope, Hitachi, Japan) with a 15 kV beam.

The printed samples on presintered NiO-YSZ substrates after firing at 800, 1000, and 1300°C were cut, embedded in epoxy and polished for SEM investigations of cross sections. Polished surfaces of all samples were coated with 10-nm carbon to increase the surface conductivity. The samples sintered at 800 and 1000°C were investigated by a Hitachi Tabletop TM3000 SEM (15 kV charge reduction mode), using the equipped high-sensitivity semiconductor detector.

The sample sintered at 1300°C was investigated by a Zeiss Merlin SEM with a field emission gun operating at 10 kV. A high-efficiency secondary electron (SE2) detector and an energy selective backscattered detector (BSE) were used simultaneously to record images at each selected area. Besides the polished cross section, the surface of the printed GDC layer (top view) was also investigated. EDS analysis was conducted on the polished cross-section by the equipped Bruker XFlash 6 EDS detector with a standard energy resolution 0.129 keV. An accelerating voltage of 10 kV was used giving a theoretical lateral resolution around 0.5 μm, and a working distance 8 mm was used. The Esprit 1.9 software was used to record and process the spectra by the 'Hyper Map' method. The total spectrum was background subtracted and deconvoluted.

## RESULTS AND DISCUSSION

### Continuous hydrothermal flow synthesis (CHFS) of Gd-doped CeO2 nanoparticles

Figure 2 presents XRD patterns of the synthesized particles. All diffraction peaks of particles synthesized at conditions with various pH values matched to those of $CeO_2$ (cubic, Fm-3 m, ICSD *PDF* 01-081-0792). This indicates the presence of cubic $CeO_2$ phase in all particles, although the shift in 2h positions of experimental patterns compared with the standard shows the difference in cell parameters arising from $Gd^{3+}$ dopants. The peak broadening suggests that nanoparticles were obtained. The crystallite size was estimated from the FWHM of peaks, and the sizes estimated from the first three peaks corresponding to (111), (200) and (220) planes of a $CeO_2$ crystal are summarized in Table 1 (more details in Supporting information). Depending on the experimental conditions, crystallite sizes between 6 and 30 nm can be obtained. The crystallite size first decreased with increasing pH values (pH 1.9-5.5) and then increased when more alkali was added (pH 9-13.1).

The addition of alkali solution (KOH in this study) significantly affected the morphology and size of particles crystallized from the solution in a CHFS process. When the synthesis was conducted in the absence of KOH, well-crystallized particles with a fully developed octahedral shape displaying flat surfaces and sharp corners were obtained (Figure 3). Depending on the view direction in the TEM, projections with different geometries were observed, e.g. square projections viewed in [001] direction (Figure 3A), and rhombohedral projections in [110] direction (Figure 3B and 3C). FFT (insets in Figure 3B and 3C) conversions show that these octahedral were enclosed by {111} facets (flat surfaces), whereas {002} facets disappeared (sharp corners). As summarized in Table 1, the edge length size of these single-crystalline octahedral was 39-45 nm measured in TEM. However, crystallite size estimated from XRD was relatively smaller and varied among peaks (Table 1 and S1), indicating a pronounced anisotropic growth of crystallites.

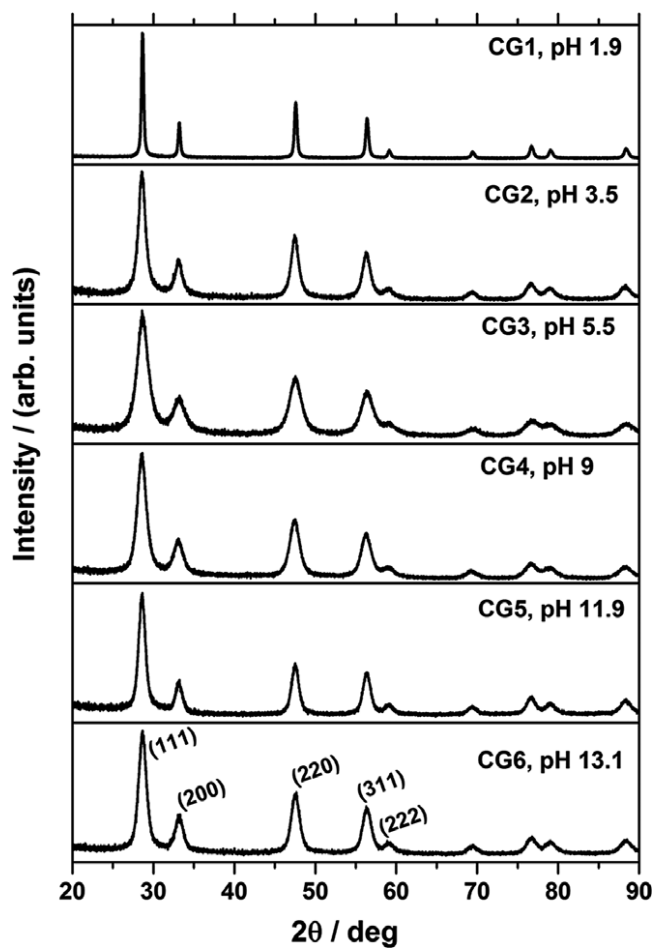

**FIGURE 2**: XRD patterns of $Gd_xCe_{1-x}O_{2-d}$ particles prepared by CHFS

In comparison, when KOH was introduced to the syntheses (CG2, 3, 4, 5, 6, and CG-s in Table 1), the obtained particles were very different with respect to the size and morphology, as presented in Figure 4. KOH was added in order to tune the pH values during the syntheses, which were measured from the final slurries of particles. Firstly, the particle size markedly decreased when the alkali was involved in the syntheses for all samples (compared with that of CG1). Moreover, in acidic conditions, the particle size decreased from 8(1) nm (Figure 4A) to 6(1) nm (Figure 4B) as the amount of KOH was increased (pH changed from 3.5 to 5.5). In base conditions, however, the particle size slightly increased from 7(1) nm (Figure 4C) to 8(2) nm (Figure 4E) as the amount of KOH was increased (pH changed from 9 to 11.9). For all samples synthesized with KOH, the crystallite sizes estimated from different peaks in XRD patterns were relatively close to each other (Table 1 and S1). The particle size statistically quantified with TEM images was also close to the crystallite size for these samples.

In terms of the morphology, all these particles displayed polyhedral shapes (Figure 4), while some of them had relatively flat surfaces (red outlines), others were more spherical-like (blue outlines). As shown in insets, particles displaying flat surfaces were enclosed by {111} and {002} planes, which in some other publications[44,45] have been referred as 'truncated octahedra'. On adding more KOH during syntheses, the amount of particles with flat surfaces decreased while increasing amounts of spherical-like particles were obtained. For instance, a particle (CG6, pH = 13.1) enclosed by {111} and {002} planes (inset in Figure 4E) appeared more spherical like.

In the absence of surfactants or templates, the morphology of particles mostly depends on crystallographic structures that in turn affect the growth behavior. $CeO_2$-based oxide has a face-centered cubic (FCC) structure, and the surface energy of low-index planes follows $c_{\{111\}} < c_{\{200\}} < c_{\{110\}}$.[9,44] Therefore, the growth rate of {111} planes is the lowest and $CeO_2$ octahedra terminated by {111} planes have the lowest surface energy, i.e. the highest stability. In a typical hydrothermal process, the formation of particles follows a dissolution-nucleation-growth mechanism. $CeO_2$ particles crystallized from the solution tend to be in an octahedral shape with a minimized surface energy. This was observed in all samples, displaying as fully developed octahedra with flat surfaces and sharp corners (Figure 3) or as polyhedral ('truncated octahedra') either with flat surfaces or with 'spherical-like' appearances (Figure 4). Wang et al.[44] found a 'size effect' that the polyhedra dominated in particles with a size 3-10 nm as {002} planes had not disappeared completely. Here, the observations are similar, as polyhedra were dominant for particles between 6 and 8 nm (Figure 4A-E) while fully developed octahedra dominated in large particles (Figure 3).

The resulting particle size reflects the combined effect of the nucleation and the growth processes. The hydrothermal synthesis of $CeO_2$ from $Ce^{3+}$ starts from the hydration between $Ce^{3+}$ and $OH^-$ groups either from alkaline addition or self-ionization of water, followed by formation of $Ce(OH)_3$ nuclei, oxidation and growth of $CeO_2$ crystals[46]. One can expect a high concentration of $Ce^{3+}$ and a high solubility of newly formed $Ce(OH)_3$ intermediates (by reaction between $Ce^{3+}$ and $OH^-$ from water) in the very acidic solution during the synthesis of CG1 where no KOH was added (Table 1). Therefore, the amount of nuclei was relatively low whereas a comparatively rapid growth of nuclei was facilitated by the fast mass diffusion in the solution, favoring large crystals.[9] In comparison, during syntheses of CG2 to CG6 (Table 1), KOH was introduced and more $Ce(OH)_3$ intermediates were formed in the precursor flow before heating by $scH_2O$. The solubility of $Ce(OH)_3$ was much lower in the solution particularly in more alkaline conditions (with more KOH, CG4 to CG6). Compared with the synthesis of CG1, a higher degree of supersaturation was achieved, and as a result, the driving force for nucleation was larger, facilitating the formation of more numerous and smaller crystals.[9]

The effect of pH on the composition of the synthesized Gd-doped $CeO_2$ particles is shown in Figure 5. The molar ratio of Ce to Gd was carefully controlled at 80:20 in the reactant solutions, however the ratio of Ce to Gd in particles deviated as Gd deficiency was detected in particles synthesized at acidic conditions (CG1, CG2 and CG3). This suggests that in order to get a stoichiometry composition in Gd-doped $CeO_2$ particles the pH should be controlled, preferably in the alkaline range. A similar result was observed in CHFS of Y-doped $ZrO_2$ nanoparticles where the pH value had to be >8 to achieve a comparable conversion rate of Y and Zr.[47]

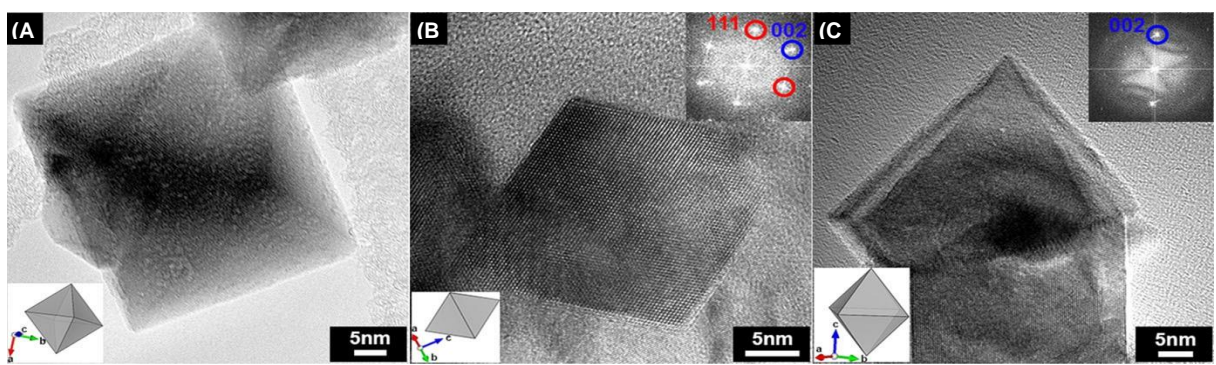

**FIGURE 3**: BF-TEM images of CG1 nanoparticles displaying an octahedron morphology with flat surfaces and sharp corners. A, Particle viewed in [001] direction; both (B) and (C) are particles viewed in [110] direction, but oriented in *c*-axis; the insets are live FFT conversions, showing families of {111} and {002} planes.

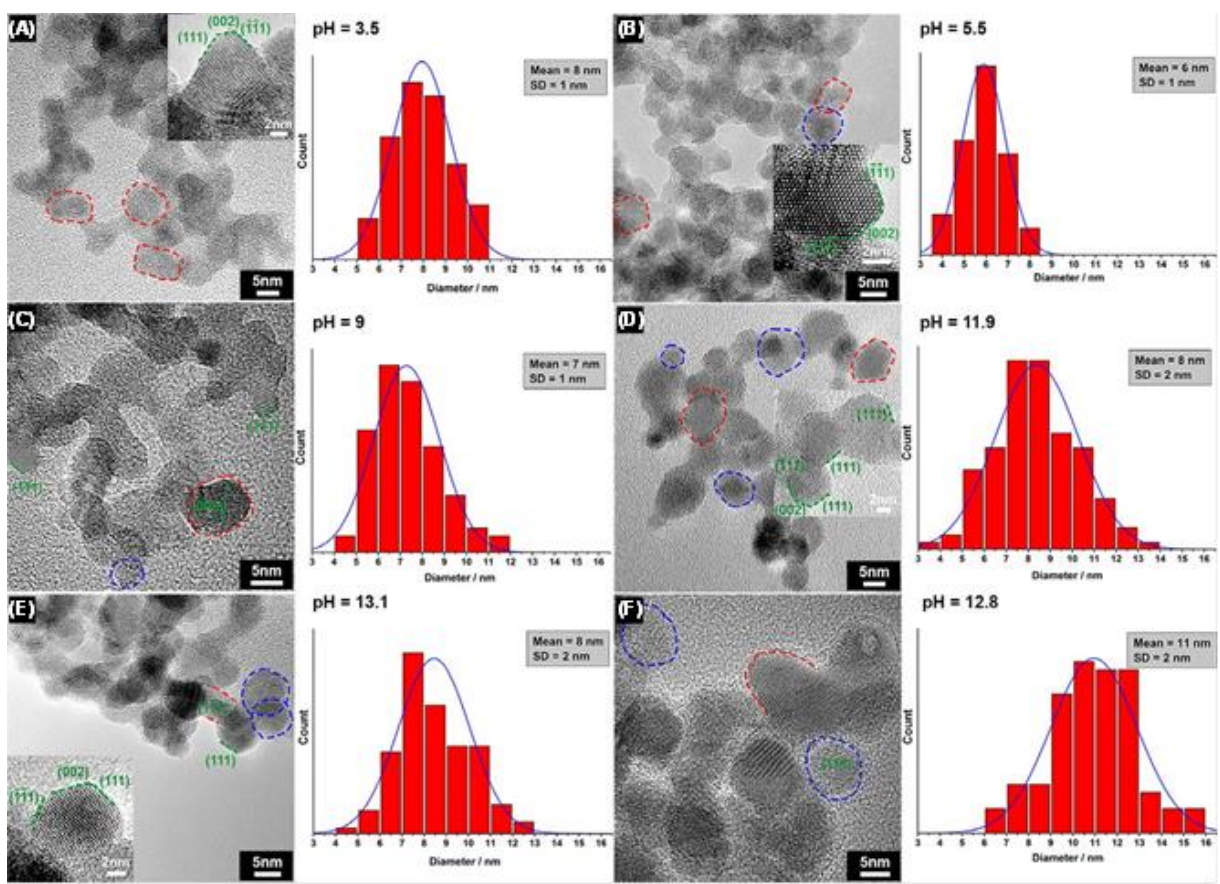

**FIGURE 4**: BF-TEM images of GDC nanoparticles and particle size distribution. A, CG2, pH = 3.5; the inset shows a HR-TEM image of a GDC nanoparticle with the truncated octahedron morphology. B, CG3, pH = 5.5; the inset shows a HR-TEM image of a GDC nanoparticle with the truncated octahedron morphology. C, CG4, pH = 9. D, CG5, pH = 11.9; the inset shows a HR-TEM image of GDC nanoparticles with relatively spherical morphology and with truncated octahedron morphology. E, CG6, pH = 13.1; the inset shows a HR-TEM image of a GDC nanoparticle enclosed by (002) and (111) planes however is spherical-like. F, CG-s, pH = 12.8. The red dashed curves and the blue dashed curves outline the particles with flat surfaces and relatively spherical particles separately.

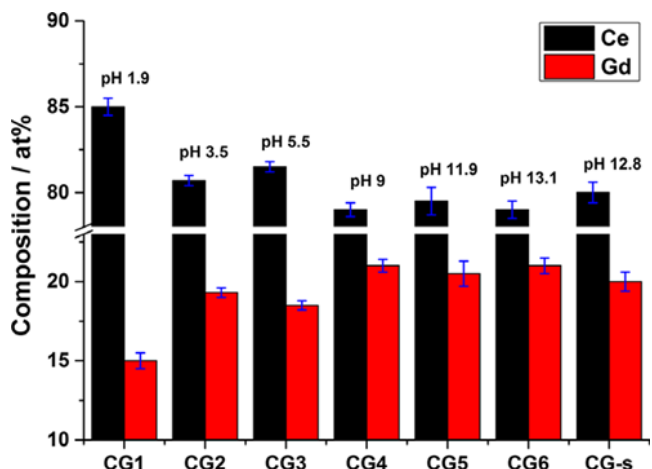

**FIGURE 5**: Chemical compositions in atomic percent (at %) of $Gd_xCe_{1-x}O_{2-d}$ nanoparticles. Syntheses of all samples started from nitrate precursors composed of 80% Ce and 20% Gd in molar percent

**Properties of the formulated $Gd_{0.2}Ce_{0.8}O_{2-d}$ inks**

An up-scaled synthesis of GDC nanoparticles (CG-s) was conducted to get sufficient amounts of particles for making inks. The BET-specific surface area of the particles was measured as 91.1 m$^2$ g$^{-1}$. The particles had a relatively spherical morphology with a narrow size distribution 11 ± 2 nm (Figure 4F). Similar conversion rates of Ce and Gd were achieved, as the chemical composition of the synthesized particles ($Gd_{0.2}Ce_{0.8}O_{2-d}$, Figure 5, CG-s) was the same as the composition (Gd:Ce molar ratio 20:80) of the precursor.

The mass fraction of $Gd_{0.2}Ce_{0.8}O_{2-d}$ in the primary dispersion was determined by thermogravimetric analysis (TGA) to be 7.2 wt%. The viscosity of the primary dispersion is shown in Figure S1.

Of the ink formulated for printing on green NiO-GDC substrates, the surface tension, viscosity, and density were measured to be 27 mN m$^{-1}$, 2.1 mPa s, and 1.08 g cm$^{-3}$, respectively (Figure S2). Therefore, the calculated *Oh$^{-1}$* Number was 12.0, indicating printability.[48]

The formulated ink for printing on presintered NiO-YSZ substrates was characterized by measuring surface tension and viscosity and values of 25 mN m$^{-1}$ and 1.0 mPa s (Figure S3) were obtained, respectively. The calculated *Oh$^{-1}$* Number was 23.8, which should lead to the formation of satellite droplets according to Jang[48] and Derby.[49] Nonetheless, we observed the generation of stable single droplets during our experiments with a volume between 7 and 8 pL (Figure 6). Thus, probably the presence of nanoparticles influences the jetting behavior of the fluid extending the range of properties at which the ink results printable.

**Microstructure of the printed GDC films on substrates**

*Microstructure of GDC printed on green NiO-GDC substrates*

Optical images (Figure 7) reveal that cracking was present in the unsintered 10- and 20-layered films, but not in the 1- or 5-layered films. Therefore, it is evident that the cracking occurred during the drying after the printing process, and depended on the thickness of the printed layers, i.e., the critical cracking thickness. This is due to the small nominal particle radius (11 ± 2 nm), which affects the critical cracking thickness (proportional by the power 5/3).[50] The single-layered film sintered at 1300°C is shown in Figure 8, and micrographs of the 5-layered film are shown in Figure S4. In neither case, the obtained films were continuous, and the NiO-GDC substrate is still visible after sintering in some places (Figure 8). These results indicate that under the conditions applied it was not possible to obtain a dense film completely covering the substrate, which could function as a SOFC electrolyte. Further improvements could be by optimizing the ink formulation aiming for less particle agglomerations and a crack-free film after printing or by optimizing sintering profiles to obtain a dense GDC film.

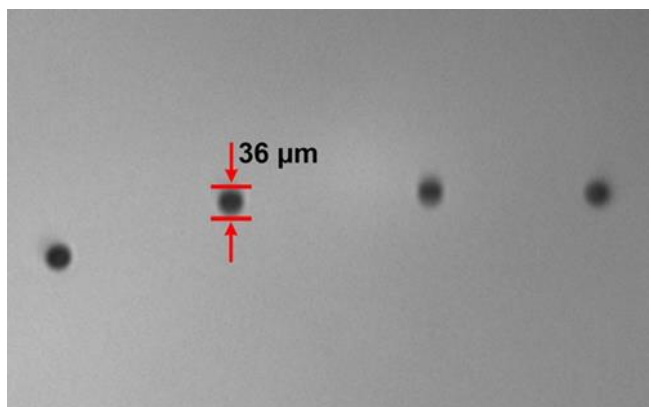

**FIGURE 6**: Droplets generated by the printhead. Despite a high value of the $Oh^{-1}$ Number, single round-shaped droplets were formed. No satellite droplets were observed.

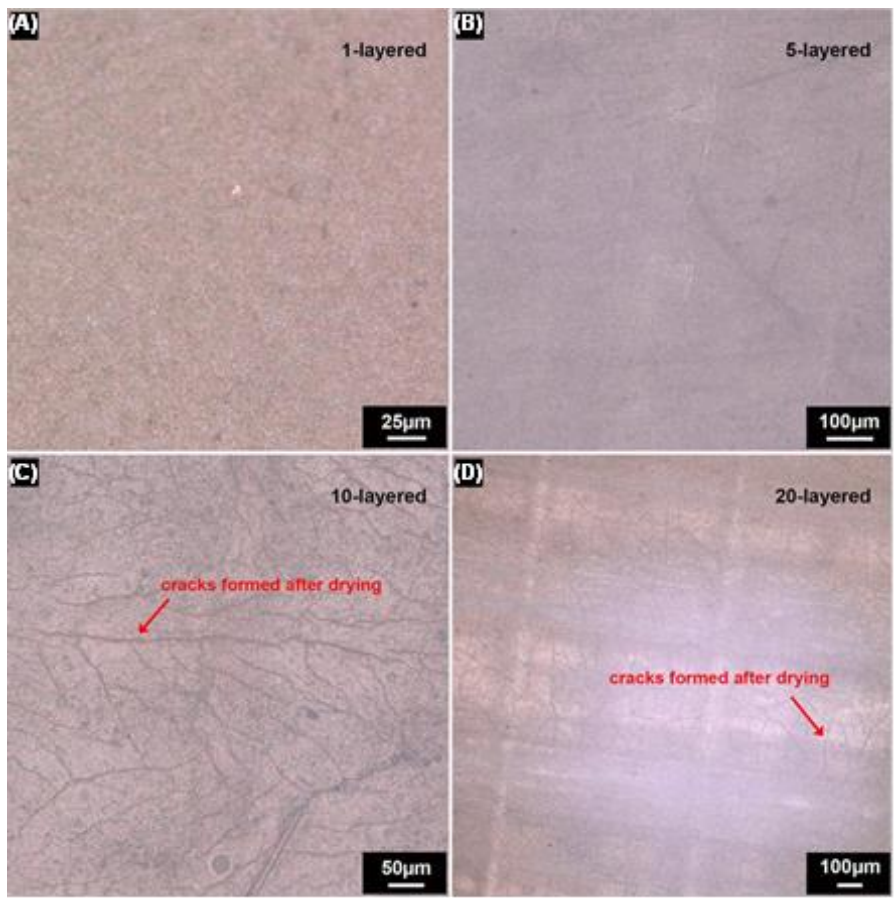

**FIGURE 7**: Optical images of the surface (top-down view) of the unsintered GDC films: A, 1 layer, (B) 5 layers, (C) 10 layers, and (D) 20 layers. Cracking can be observed in (C) and (D). Printed on a green NiO-GDC substrate

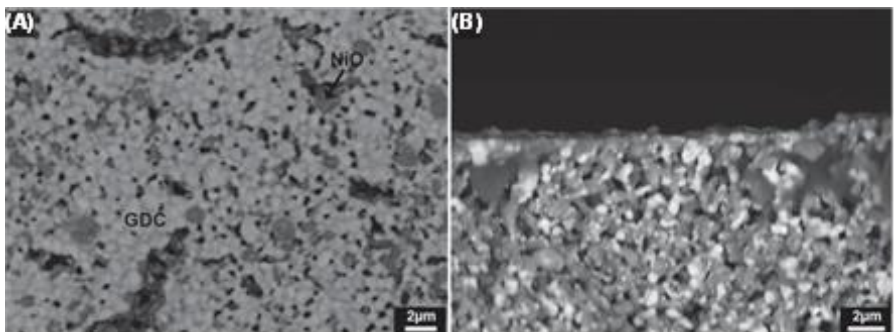

**FIGURE 8**: SEM micrographs of (A) surface view and (B) side view of the single-layered GDC film after sintering at 1300°C. Printed on a green NiO-GDC substrate.

### Microstructure of GDC printed on pre-sintered NiO-YSZ substrates

Gd-doped $CeO_2$ films were also printed on presintered NiO-YSZ substrates and sintered at 800, 1000, and 1300°C. Figure 9A presents an overview of the surface (BSE micrographs, top-down view) of the sample sintered at 1300°C displaying the following regions of interest: (i) the inkjet-printed GDC film, (ii) the transition area between the inkjet-printed GDC film and the bare substrate, and(iii) the uncovered NiO-YSZ substrate. Based on contrast, only one phase (bright gray) was observed in the printed GDC film and in part of the transition area, whereas two phases were present on the uncovered substrate's surface. The SE2 micrograph of the GDC film (Figure 9B) shows that surface structure can be described as islands of GDC grains, which are disconnected by 'trenches'. However, it is worth noticing that inside these trenches no secondary phase was present, as shown by the BSE micrographs (Figure 9C and Figure 9D). For reference, a BSE micrograph (Figure 9E) of the surface of the uncovered substrate shows NiO and YSZ differentiated by the contrast. Considering that GDC was printed on the surface of the substrate, this indicates that GDC completely covered the substrate after sintering although the shrinkage resulted in trenches and disconnected GDC islands on the top.

For further verification, side-viewed cross section of the part with the printed GDC film was investigated (Figure 10, S5, S6). The SE2 micrograph (Figure 10A) shows the 'trenches' between the islands of GDC grains on top of the printed film after sintering. The BSE micrograph (Figure 10B) shows that at the bottom of the trench as well as below the islands of GDC grains a fully dense, continuous layer was found (Figure 10A). This layer differed significantly in contrast (bright gray) compared to the NiO (dark gray). Element maps (Figure 10C-F) show that the continuous layer could be divided into two parts. The top part was based on GDC (Figure 10E) and was as thin as 0.5 μm below the trenches), whereas below this part a thin (0.8-1 μm) layer of YSZ was formed. This suggests a full coverage of the substrate by the printed GDC, which is consistent with the top-down-view observation (Figure 9D). Interestingly, the formation of the continuous YSZ layer must have occurred during the sintering at 1300°C. As shown in Figure S7D, such a layer was not present at the surface of the bare substrate in the unprinted area after sintering. In these uncovered areas, clearly NiO particles were found at the surface. From an application point of view, this phenomenon is highly advantageous, as the formed dense GDC/YSZ layers can function as a dense electrolyte with a GDC barrier layer. Fast interdiffusion between ceria and zirconia, with the formation of $ZrO_2$-$CeO_2$ solid solution, was also expected at the electrolyte-barrier layer interface. However, chemical analysis in Figure 10 shows a rather sharp chemical front at the interface. Considering the low thickness of only 1-2 μm, better performances of cells with such a microstructure can be expected compared to state-of-the-art cells with electrolyte thicknesses of around 10 μm. Importantly, the corrugated electrolyte surface (disconnected GDC islands, Figure 10A, 10B) is expected to improve the adhesion between the cathode and electrolyte, while decreasing the interfacial electrolyte-cathode resistance. The formation of 'disconnected GDC islands' could be a result of different shrinkages of the substrate and the printed layer. Further work to optimize the microstructure (e.g. by optimizing the sintering profile) and to identify the effect of the microstructure on the electrochemical performance need to be carried out.

For reference, the side-viewed cross-section micrographs of the transition area and of the uncovered substrate are presented in Figure S7. Printed GDC films in samples sintered at 800°C and 1000°C were not fully dense. Neither the continuous layer of GDC nor YSZ was observed which means that the substrate was not fully covered after sintering (Figure S8, S9).

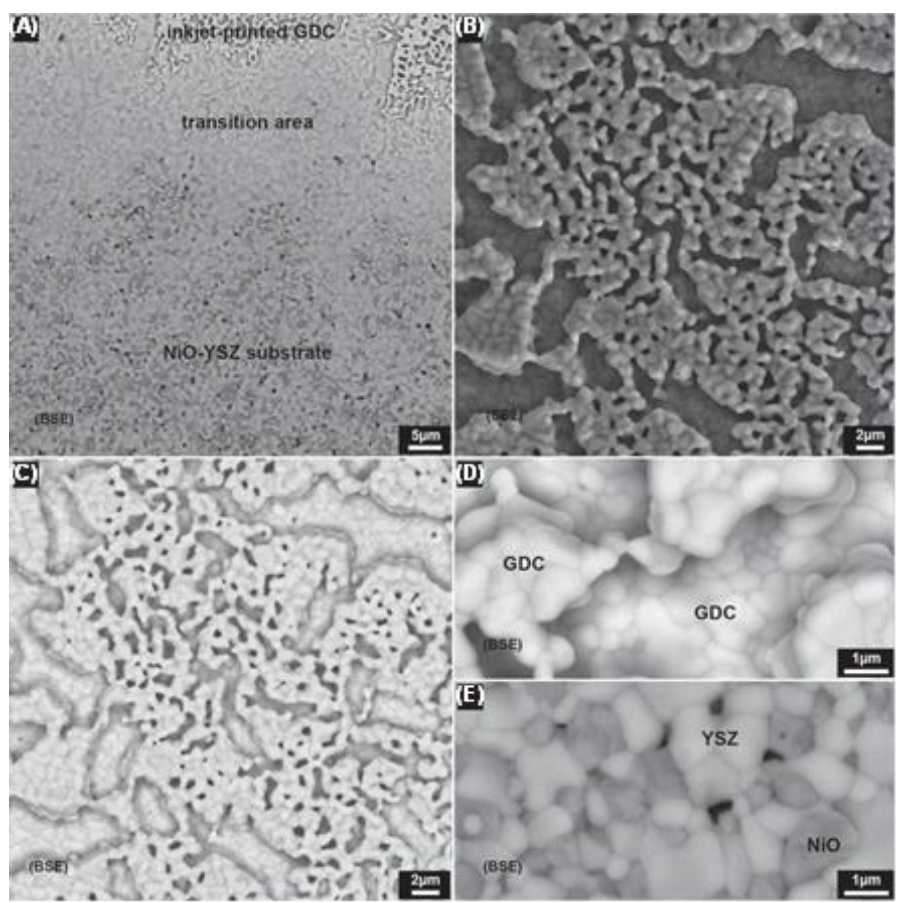

**FIGURE 9**: Top-down-view surface BSE micrograph (A) of the sample sintered at 1300°C, including the inkjet-printed GDC film, transition area and NiO-YSZ substrate. SE2 micrograph (B) and BSE micrograph (C) of the surface of the inkjet-printed GDC film. Zoom-in BSE micrographs of the surface of the GDC film (D) and of the substrate (E). All images were recorded with an acceleration voltage 15 kV.

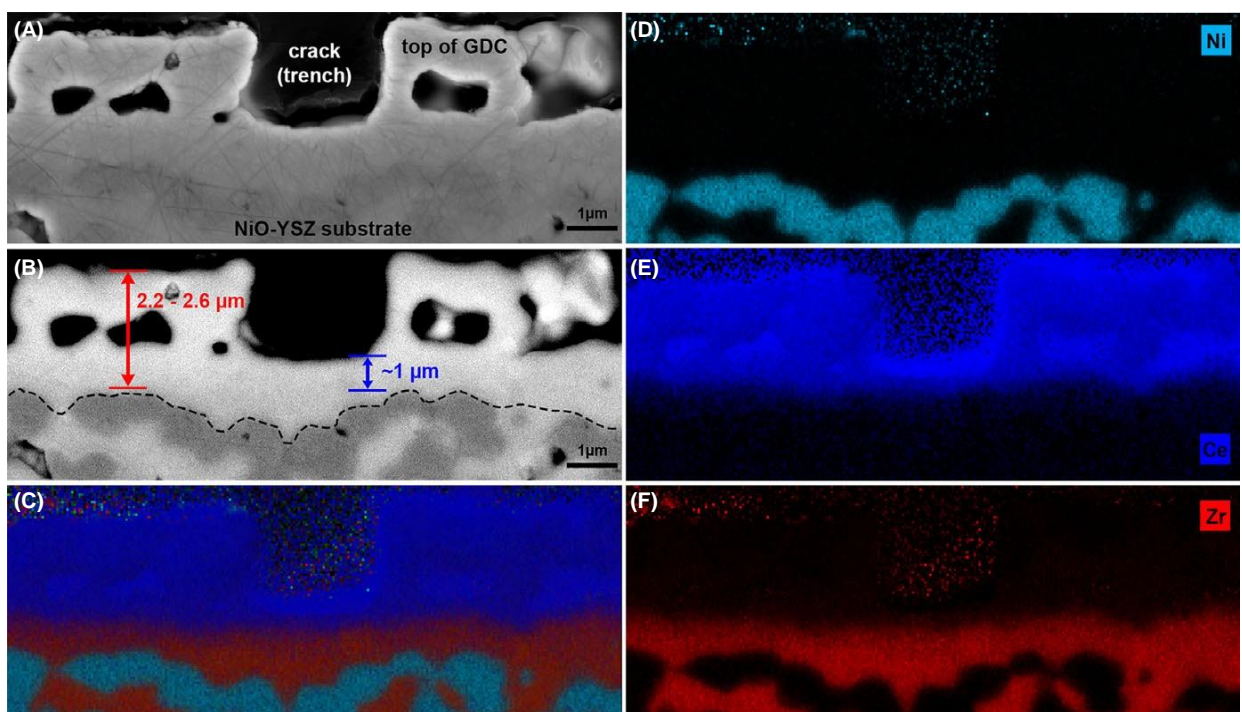

**FIGURE 10**: Side-viewed cross-section SE2 (A) and BSE (B) micrographs of the sample sintered at 1300°C. EDS element mapping of the area within the image (B): map of all elements (C), of Ni (D), Ce (E) and Zr (F). Element maps were obtained by deconvoluting the total EDS spectrum of the mapping area, working voltage 10 kV.

## CONCLUSION

Tailored $Gd_xCe_{1-x}O_{2-d}$ (GDC) nanoparticles were synthesized, using continuous hydrothermal flow synthesis. By varying the pH, the size, morphology, and composition of the GDC particles could be tailored. Here, particle sizes between 6 nm and 40 nm were obtained, while the morphology was polyhedral (flat surfaces and spherical-like) for small particles and octahedral for larger ones. A similar conversion rate of Ce and Gd, and consequently a stoichiometric composition, was achieved when alkaline conditions were applied.

The obtained $Gd_{0.2}Ce_{0.8}O_{2-d}$ particles were further processed into inks for inkjet printing. Despite the small particle size/large surface area, inks with the surface tension and the viscosity of 27 mN m$^{-1}$, 2.1 mPa s and of 25 mN m$^{-1}$, 1.0 mPa s, separately were obtained. These inks showed excellent printing behavior.

GDC layers were printed on (i) green NiO-GDC substrates and, on (ii) presintered NiO-YSZ substrates. While for the green NiO-GDC substrates, no dense films could be obtained due to cracking of the printed films after drying (10- and 20-layered) and after sintering (1- and 5-layered), the sample with 10-layered GDC film printed on NiO-YSZ substrates formed a dense continuous layer after firing at 1300°C. This continuous layer consisted of a thin (0.8-1 µm) YSZ layer covered by a GDC layer (0.5 µm) decorated with GDC islands. Such a surface structure is expected to have several advantages over state-of-the-art SOFC cells. Firstly, the thin electrolyte (here 1-2 µm) is expected to cause lower performance losses. Secondly, the disconnected islands of the printed GDC layer could provide a high number of catalytic active sides and could provide a strong bonding to the applied cathode.

## ACKNOWLEDGMENT

The authors acknowledge the ProEco project (www.proeco.dk) funded by the Danish Council for Independent Research (DFF 1335-00138) and the Cell3Ditor project (http://www.cell3ditor.eu) funded by the Fuel Cells and Hydrogen 2 Joint Undertaking (No. 700266). EU's Horizon 2020 research and innovation program, Hydrogen Europe and N.ERGHY are acknowledged for supporting the Joint Undertaking.